\documentclass[11pt,reqno,oneside]{amsart}
\usepackage{verbatim,amsmath,amssymb,cite,epsfig}
\title{Statistical Mechanics of Spatial Evolutionary Games} 
\author{Jacek Mi\c{e}kisz \\ Institute of Applied Mathematics \\
and Mechanics \\ Warsaw University  \\ ul. Banacha 2  \\ 02-097
Warsaw, Poland 
\\ e-mail: miekisz@mimuw.edu.pl} 
\begin{document} 
\baselineskip=20pt
\maketitle 

\noindent {\bf Abstract}: We discuss the long-run behavior of stochastic dynamics
of many interacting players in spatial evolutionary games. In particular, we investigate the effect 
of the number of players and the noise level on the stochastic stability of Nash equilibria. 
We discuss similarities and differences between systems of interacting players maximizing 
their individual payoffs and particles minimizing their interaction energy. 
We use concepts and techniques of statistical mechanics to study game-theoretic models.
In order to obtain results in the case of the so-called potential games, 
we analyze the thermodynamic limit of the appropriate models of interacting particles.
\vspace{3mm}

\noindent  PACS: 05.20.-y, 05.50.+q
\vspace{3mm}

\noindent Keywords: evolutionary game theory, spatial games, Nash equilibria, Gibbs states, 
stochastic stability. 

\eject

\section{Introduction}
\numberwithin{equation}{section}

Many socio-economic systems and biological populations can be modeled 
as systems of interacting individuals \cite{santa,young2,young3,nowsig,econo}. 
Here we will consider game-theoretic models of many interacting players \cite{wei,hof2,ams}. 
In such models, individuals have at their disposal certain strategies and their payoffs 
in a game depend on strategies chosen both by them and by their opponents. 
In spatial games, players are located on vertices of certain graphs and they interact 
only with their neighbors \cite{blume1,ellis1,young2,ellis2,nowak1,nowak2,linnor,doebeli,
sabo,hauert}. The central concept in game theory is that of a {\em Nash equilibrium}. 
It is an assignment of strategies to players such that no player, for fixed strategies 
of his opponents, has an incentive to deviate from his curent strategy; 
the change can only diminish his payoff.

The notion of a Nash equilibrium (called a Nash configuration in spatial games) 
is similar to the notion of a ground-state configuration 
in classical lattice-gas models of interacting particles. We will discuss similarities 
and differences between systems of interacting players maximizing their individual payoffs 
and particles minimizing their interaction energy. 

One of the fundamental problems in game theory is the equilibrium selection in games
with multiple Nash equilibria. One of the selection methods 
is to construct an appropriate dynamical system 
where in the long run only one equilibrium is played with a high frequency. 
Here we will discuss a stochastic adaptation dynamics of a population 
with a fixed number of players. In discrete moments of times, 
players adapt to their neighbors by choosing with a high probability
the strategy which is the best response, i.e. the one which maximizes 
the sum of the payoffs obtained from individual games. With a small probability, representing 
the noise of the system, they make mistakes. To describe the long-run behavior 
of such stochastic dynamics, Foster and Young \cite{foya} introduced 
a concept of stochastic stability. A configuration of the system 
(an assignment of strategies to lattice sites in spatial games) is 
{\em stochastically stable} if it has a positive probability in the stationary state 
of the above dynamics in the zero-noise limit, that is zero probability of mistakes. 
It means that in the long run we observe it with a positive frequency.
However, for any arbitrarily low but fixed noise, if the number 
of players is big enough, the probability of any individual configuration 
is practically zero. It means that for a large number of players, 
to observe a stochastically stable configuration we must assume that players 
make mistakes with extremely small probabilities. On the other hand, it may happen that 
in the long run, for a low but fixed noise and sufficiently big number 
of players, the stationary state is highly concentrated on an ensemble
consisting of one Nash configuration and its small perturbations, 
i.e. configurations, where most players play the same strategy.
We will call such configurations {\em ensemble stable.}
We will show that these two stability concepts do not necessarily coincide.

In the so-called potential games, for any given configuration, 
payoffs of all players are the same \cite{mon}. 
Such systems are therefore analogous to those of interacting particles, 
where instead of maximizing payoffs, particles minimize their interaction energy. 
Stationary states of the stochastic dynamics with the Boltzmann-type updating
are then finite-volume Gibbs distributions describing an equilibrium behavior
of corresponding systems of interacting particles in the grand-canonical ensemble. 
We use techniques and results of statistical mechanics to describe the long-run behavior 
of potential games. We investigate a thermodynamic limit, i.e. the limit of the infinite 
number of players. 
  
We will present examples of spatial games with three strategies 
where concepts of stochastic stability and ensemble stability do not coincide. 
In particular, we may have the situation, where a stochastically stable strategy 
is played in the long run with an arbitrarily low frequency. 

We will also discuss briefly nonpotential games. Stationary states of such games 
cannot be explicitly constructed as before. We must therefore resort to different methods. 
We will use a tree characterization of stationary states \cite{freiwen1,freiwen2}.

In Section 2, we introduce spatial games with local interactions. 
In Section 3, we present stochastic dynamics and the concept of stochastic stability 
of Nash configurations. In Section 4, we introduce our concept of ensemble stability
and present examples of games where stochastically stable
Nash configurations are played in the long run with arbitrarily 
small probabilities if the noise level is low and the number of players 
is big enough. We will also discuss an effect of adding a dominated
strategy to a game with two strategies. In particular, the presence of such a strategy
may cause a stochastically stable strategy to be observed in the long run 
with a frequency close to zero. In Section 5, we discuss the long-run behavior 
of a certain example of a nonpotential game. Discussion follows in Section 6.

\newtheorem{theo}{Theorem}
\newtheorem{defi}{Definition}
\newtheorem{hypo}{Hypothesis}
\newtheorem{prop}{Proposition}
\section{Spatial Games with Local Interactions}

In order to characterize a game-theoretic model, one has to specify the set of players, 
strategies they have at their disposal and payoffs they receive.  
Here we will discuss only two-player games with two or three pure strategies.
In addition, players may use mixed strategies. A mixed strategy is a probability 
distribution on the set of pure strategies. We begin with games with two pure strategies 
and two symmetric Nash equilibria. A generic payoff matrix is given by
\newpage

\noindent {\bf Example 1}
\vspace{2mm}

\hspace{23mm} A  \hspace{2mm} B   

\hspace{15mm} A \hspace{3mm} a  \hspace{3mm} b 

U = \hspace{6mm} 

\hspace{15mm} B \hspace{3mm} c  \hspace{3mm} d,

where the $ij$ entry, $i,j = A, B$, is the payoff of the first (row) player when
he plays the strategy $i$ and the second (column) player plays the strategy $j$. 
We assume that both players are the same and hence payoffs of the column player are given 
by the matrix transposed to $U$; such games are called symmetric. 
Let $(x,1-x)$ be a mixed strategy, where $x$ is the probability of playing 
$A$ and $1-x$ of playing $B$. We then assume that the payoff received by a player 
using a mixed strategy $(x,1-x)$ against a player using $(y,1-y)$ is the average (expected) 
payoff given by $x[ay+b(1-y)]+(1-x)[(cy+d(1-y)]$.    

An assignment of strategies to both players is a {\em Nash equilibrium}, if for each player, 
for a fixed strategy of his opponent, changing the current strategy will not increase 
his payoff.

We will discuss games with multiple Nash equilibria. 
If $a>c$ and $d>b$, then both $(A,A)$ and $(B,B)$ are Nash equilibria. 
If $a+b<c+d$, then the strategy $B$ has a higher expected payoff against a player playing 
both strategies with the probability $1/2$. We say that $B$ {\em risk dominates} the strategy $A$ 
(the notion of the risk-dominance was introduced and thoroughly studied 
by Hars\'{a}nyi and Selten \cite{hs}). If at the same time $a>d$, 
then we have a selection problem of choosing between the payoff-dominant 
(Pareto-efficient) equilibrium $(A,A)$ and the risk-dominant $(B,B)$.

We will study populations with a finite number of individuals
playing two-player games. In spatial games, players occupy sites of certain lattices and 
interact only with their neighbors. 

Let $\Lambda$ be a finite subset of the simple lattice ${\bf Z}^{2}$ 
(for simplicity of presentation we assume periodic boundary conditions,
i.e. we place players on a two-dimensional torus). Every site of $\Lambda$ 
is occupied by one player who has at his disposal one of $k$ different pure strategies 
(player do not use mixed strategies). Let $S$ be the set of pure strategies, 
then $\Omega_{\Lambda}=S^{\Lambda}$ is the set of all configurations of players, 
that is all possible assignments of strategies to individual players. For every $i \in \Lambda$, 
$X_{i}$ is the strategy of the $i-$th player in the configuration $X \in \Omega_{\Lambda}$ 
and $X_{-i}$ denotes strategies of all remaining players; $X$ therefore can be represented 
as the pair $(X_{i},X_{-i})$. Let $U: S \times S \rightarrow R$ be a matrix 
of payoffs of our game. Every player interacts only with his neighbors and his payoff 
is the sum of the payoffs resulting from individual games.
We assume that he has to use the same strategy for all neighbors. 
Let $N_{i}$ denote the neighborhood of the $i-$th player. 
For the nearest-neighbor interaction we have $N_{i}=\{j; |j-i|=1\}$,
where $|i-j|$ is the distance between $i$ and $j$.
For $X \in \Omega_{\Lambda}$ we denote by $\nu_{i}(X)$ the payoff 
of the $i-$th player in the configuration $X$:
\begin{equation}
\nu_{i}(X)=\sum_{j \in N_{i}}U(X_{i}, X_{j})
\end{equation}

\begin{defi}
$X \in \Omega_{\Lambda}$ is a {\em Nash configuration} if for every $i \in \Lambda$
and $Y_{i} \in S,$ $\nu_{i}(X_{i},X_{-i}) \geq \nu_{i}(Y_{i},X_{-i})$.
\end{defi}

In Example 1, there are two homogeneous Nash configurations, $X^{A}$ and $X^{B}$, 
in which all players play the same strategy, $A$ or $B$ respectively. 

Let us notice that the notion of a Nash configuration is similar to the notion 
of a ground-state configuration in classical lattice-gas models of interacting particles.
We have to identify agents with particles, strategies with types of particles
and instead of maximizing payoffs we should minimize interaction energies.
There are however profound differences. First of all, 
ground-state configurations can be defined only for symmetric matrices; 
an interaction energy is assigned to a pair of particles, payoffs are assigned 
to individual players. It may happen that if a player switches a strategy 
to increase his payoff, the payoff of his opponent and of the entire population decreases.
Moreover, ground-state configurations are stable with respect to all local changes,
not just one-site changes like Nash configurations.
It means that for the same symmetric matrix $U$, there may exist a configuration 
which is a Nash configuration but not a ground-state configuration 
for the interaction marix $-U$. The simplest example is given by Example 1
with $a=2, b=c=0$, and $d=1$. $X^{A}$ and $X^{B}$ are Nash configurations but only $X^{A}$ 
is a ground-state configuration for $-U.$  

For any classical lattice-gas model there exists at least one 
ground-state configuration. It may happen that a game with a nonsymmetric 
payoff matrix may not posess a Nash configuration. The classical example is that 
of the Rock-Scissors-Paper game given by the following matrix.
\vspace{2mm}

\noindent {\bf Example 2}

\hspace{23mm} R  \hspace{2mm} S \hspace{2mm} P  

\hspace{15mm} R  \hspace{3mm} 1  \hspace{3mm} 2 \hspace{3mm} 0

U = \hspace{6mm} S \hspace{3mm} 0  \hspace{3mm} 1 \hspace{3mm} 2

\hspace{15mm} P \hspace{3mm} 2  \hspace{3mm} 0 \hspace{3mm} 1

This two-player game does not have a Nash equilibrium in pure strategies. It has a unique 
mixed Nash equilibrium in which both players use a mixed strategy, playing all three
pure strategies with the probability $1/3$. 
One may show that the game does not have any Nash configurations on ${\bf Z}$ 
and ${\bf Z}^{2}$ with nearest-neighbor interactions but it has multiple Nash configurations 
on the triangular lattice.

In short, ground-state configurations minimize the total energy of a particle system, 
Nash configurations do not necessarily maximize the total payoff of a population of agents.

\section{Stochastic Stability}
We describe now the deterministic dynamics of the {\em best-response rule}. 
Namely, at each discrete moment of time $t=1,2,...$, a randomly chosen player may update 
his strategy. He simply adopts the strategy, $X_{i}^{t}$, which gives him 
the maximal total payoff $\nu_{i}(X_{i}^{t}, X^{t-1}_{-i})$ 
for given $X^{t-1}_{-i}$, a configuration of strategies 
of remaining players at time $t-1$. 

Now we allow players to make mistakes with a small probability, 
that is to say they may not choose the best response. A probability of making
a mistake may depend on the state of the system (a configuration of strategies
of neighboring players). We will assume that this probability is a decreasing function
of the payoff lost as a result of a mistake \cite{blume1}. 
In the Boltzmann-type updating (called a {\em log-linear rule} in the economics/game theory literature), 
the probability of chosing by the $i-$th player
the strategy $X_{i}^{t}$ at time $t$ is given by the following conditional probability:

\begin{equation}
p_{i}^{T}(X_{i}^{t}|X_{-i}^{t-1})=
\frac{e^{(1/T)\nu_{i}( X_{i}^{t},X_{-i}^{t-1})}}{\sum_{X_{i} \in S}
e^{(1/T)\nu_{i}(X_{i},X_{-i}^{t-1})}},
\end{equation}
where $T>0$ measures the noise level.

Let us observe that if $T \rightarrow 0$, 
$p_{i}^{T}$ converges to the best-response rule.
Our stochastic dynamics is an example of an ergodic Markov chain 
with $|S^{\Lambda}|$ states. Therefore, it has a unique stationary 
distribution (a stationary state) which we denote by $\mu_{\Lambda}^{T}.$  

The following definition was first introduced by Foster and Young \cite{foya}.

\begin{defi}
$X \in \Omega_{\Lambda}$ is {\bf stochastically stable} 
if  $\lim_{T \rightarrow 0}\mu_{\Lambda}^{T}(X) >0.$
\end{defi} 
If $X$ is stochastically stable, then the frequency of visiting $X$ converges to
a positive number along any time trajectory almost surely. It means that in 
the long run we observe $X$ with a positive frequency.
In examples below, we consider games with symmetric Nash equilibria 
and homogeneous Nash configurations. By a stochastic stability of a strategy or a Nash equilibrium
we mean a stochastic stability of the corresponding Nash configuration.  

The notion of a stochastically stable Nash configuration is analogous to the notion 
of a low-temperature stable ground-state configurations, i.e. the one which gives 
rise to a low-temperature equilibrium phase.

Stationary distributions of the Boltzmann dynamics can be explicitly constructed 
for the so-called potential games. A game is called 
a {\em potential game} if its payoff matrix can be changed 
to a symmetric one by adding payoffs to its columns. 
Such a payoff transformation does not change the strategic 
character of the game, in particular it does not change the set of its
equilibria and their stochastic stability. More formally,
it means that there exists a symmetric matrix $V$ called a potential
of the game such that for any three strategies $A, B, C \in S$
\begin{equation}
U(A,C)-U(B,C)=V(A,C)-V(B,C).
\end{equation} 

It is easy to see that every game with two strategies has a potential 
$V$ with $V(A,A)=a-c$, $V(B,B)=d-b$, and $V(A,B)=V(B,A)=0.$ 
If $V$ is a potential of the stage game, then $V(X)=\sum_{(i,j)\in \Lambda}V(X_{i},X_{j})$ 
is a potential of a configuration $X$ in the corresponding spatial game. 
The unique stationary state of a potential game with the Boltzmann dynamics 
is given by the following formula \cite{young2}:

\begin{equation}
\mu^{T}_{\Lambda}(X)=\frac{e^{(1/T)\sum_{(i,j)\in \Lambda}V(X_{i},X_{j})}}
{\sum_{Z \in \Omega_{\Lambda}}e^{(1/T)\sum_{(i,j)\in \Lambda}V(Z_{i},Z_{j})}}.
\end{equation}

$\mu^{T}_{\Lambda}$ is a so-called finite-volume Gibbs state -
a probability distribution describing an equilibrium behavior of a system 
of particles with a two-body Hamiltonian $-V$ and the temperature $T$. 
The limit $\lim_{T \rightarrow 0}\mu^{T}_{\Lambda}$ is a ground-state measure
supported by ground-state configurations, that is Nash configurations with the biggest $V$.
It follows from (3.3) that stochastically stable Nash configurations are those with
the biggest potential. In particular, in spatial games with two strategies 
and two Nash equilibria, the risk-dominant configuration $X^{A}$ 
is stochastically stable.  

In Section 4, using statistical mechanics methods, 
we will study the behavior of $\mu^{T}_{\Lambda}$ in the limit of the infinite 
number of players, i.e. in the thermodynamic limit, 
for various two-player games with three pure strategies.

\section{Ensemble Stability}

The concept of stochastic stability involves individual configurations of players. 
In the zero-noise limit, a stationary state is usually concentrated on one 
or at most few configurations. However, for a low but fixed noise and
for a big number of players, the probability of any individual configuration 
of players is practically zero. The stationary state, however, may be highly 
concentrated on an ensemble consisting of one Nash configuration and its small 
perturbations, i.e. configurations, where most players use the same strategy. 
Such configurations have relatively high probability in the stationary state. 
We call such configurations ensemble stable.

\begin{defi}
$X \in \Omega_{\Lambda}$ is {\bf $\epsilon$-ensemble stable}
if $\mu_{\Lambda}^{T}(Y \in \Omega_{\Lambda};Y_{i} \neq X_{i}) < \epsilon$
for any $i \in \Lambda$ if $\Lambda \supset \Lambda(T)$ for some $\Lambda(T)$. 
\end{defi}

\begin{defi}
$X \in \Omega_{\Lambda}$ is {\bf low-noise ensemble stable}
if for every $\epsilon>0$ there exists $T(\epsilon)$ such that if
$T<T(\epsilon)$, then $X$ is $\epsilon$-ensemble stable.
\end{defi}

If $X$ is $\epsilon$-ensemble stable with $\epsilon$ close to zero, then the ensemble 
consisting of $X$ and configurations which are different from $X$ at at most few sites has 
the probability close to one in the stationary state. It does not follow, however, 
that $X$ is necessarily low-noise ensemble or stochastically stable as it happens 
in examples presented below.
\vspace{2mm}

\noindent {\bf Example 3}
\vspace{2mm}

\noindent Players are located on a finite subset $\Lambda$ of ${\bf Z}^{2}$ 
(with periodic boundary conditions) and interact with their 
four nearest neighbors. They have at their disposal three pure strategies: 
$A, B,$ and $C$. The payoffs are given by the following symmetric matrix:
\vspace{5mm}

\hspace{23mm} A  \hspace{2mm} B \hspace{2mm} C  

\hspace{15mm} A  \hspace{2mm} 1.5  \hspace{2mm} 0 \hspace{2mm} 1

U = \hspace{6mm} B \hspace{3mm} 0  \hspace{3mm} 2 \hspace{3mm} 1

\hspace{15mm} C \hspace{3mm} 1  \hspace{3mm} 1 \hspace{3mm} 2

Our game has three Nash equilibria, $(A,A), (B,B)$, and $(C,C)$,
and the corresponding spatial game has three homogeneous Nash configurations:
$X^{A}, X^{B}$, and $X^{C}$. Let us notice that $X^{B}$ and $X^{C}$ have the maximal
payoff in every finite volume and therefore they are ground-state configurations for $-U$
and $X^{A}$ is not.

The unique stationary state of the Boltzmann dynamics (3.1) is a finite-volume Gibbs state 
and is given by (3.3) with $V$ replaced by $U$. 
$\sum_{(i,j)\in \Lambda}U(X^{k}_{i},X^{k}_{j})-\sum_{(i,j)\in \Lambda}U(Y_{i},Y_{j})>0$,
for every $Y \neq X^{B} and X^{C}$, $k=B,C$,
and $\sum_{(i,j)\in \Lambda}U(X^{B}_{i},X^{B}_{j})=\sum_{(i,j)\in \Lambda}U(X^{C}_{i},X^{C}_{j}).$
It follows that $\lim_{T \rightarrow 0}\mu_{\Lambda}^{T}(X^{k})=1/2$, 
$k=B, C $ so $X^{B}$ and $X^{C}$ are stochastically stable.
Let us investigate the long-run behavior of our system for large $\Lambda$, 
that is for a big number of players. Observe that 
$\lim_{\Lambda \rightarrow {\bf Z}^{2}}\mu_{\Lambda}^{T}(X)=0$ 
for every $X \in \Omega = S^{{\bf Z}^{2}}$.
Hence for large $\Lambda$ and $T>0$ we may only observe,
with reasonable positive frequencies, ensembles of configurations 
and not particular configurations. We will be interested in ensembles 
which consist of a Nash configuration and its small perturbations, 
that is configurations, where most players use the same strategy. 
We perform first the limit $\Lambda \rightarrow {\bf Z}^{2}$
and obtain a so-called infinite-volume Gibbs state in the temperature $T,$
\begin{equation}
\mu^{T} = \lim_{\Lambda \rightarrow {\bf Z}^{2}}\mu_{\Lambda}^{T}.
\end{equation}

It describes, in the thermodynamic limit, the equilibrium behavior of a system 
of interacting particles. Equilibrium behavior of such system results from the competition between
its energy $U$ and entropy $S$, i.e. the minimization of their free energy $F=U-TS$. We will show that
it is the entropy which is responsible for the ensemble stability of some Nash configurations (ground-state
configurations) in the limit of the infinite number of players (lattice sites). The phase transition 
of the first kind is manifested by the existence of multiple Gibbs states for a given Hamiltonian 
and temperature.

In order to investigate the stationary state of our example, we will apply a technique developed 
by Bricmont and Slawny \cite{brsl1,brsl2}. They studied low-temperature stability of the so-called dominant 
ground-state configurations. It follows from their results that

\begin{equation}
\mu^{T}(X_{i}=C)>1-\epsilon(T)  
\end{equation}
for any $i \in {\bf Z}^{2}$ and $\epsilon(T) \rightarrow 0$ as $T \rightarrow 0$. 

We will recall in Appendix A their proof adapted to our model.
The following theorem is a simple consequence of (4.2).
\begin{theo}
$X^{C}$ is low-noise ensemble stable.
\end{theo} 

We see that for any low but fixed $T$, if the number of players is big enough,
then in the long run, almost all players use $C$ strategy. 
On the other hand, if for any fixed number of players, $T$ is lowered substantially,
then all three strategies appear with frequencies close to $1/2$.

Let us sketch briefly the reason of such a behavior.
While it is true that both $X^{B}$ and $X^{C}$ have the same potential 
which is the half of the payoff of the whole system (it plays the role 
of the total energy of a system of interacting particles), 
the $X^{C}$ Nash configuration has more lowest-cost excitations. 
Namely, one player can change its strategy and switch to either
$A$ or $B$ and the total payoff will decrease by $8$ units. Players 
in the $X^{B}$ Nash configuration have only one possibility, 
that is to switch to $C$; switching to $A$ decreases the total payoff by $16$. 
Now, the probability of the occurrence of any configuration in the Gibbs state 
(which is the stationary state of our stochastic dynamics) 
depends on the total payoff in an exponential way. 
One can prove that the probability of the ensemble consisting of the $X^{C}$ Nash 
configuration and configurations which are different from it 
at few sites only is much bigger than the probability of the analogous
$X^{B}$-ensemble. It follows from the fact that the $X^{C}$-ensemble 
has many more configurations than the $X^{B}$-ensemble. On the other hand,
configurations which are outside $X^{B}$ and $X^{C}$-ensembles 
appear with exponentially small probabilities. It means that for large enough systems 
(and small but not extremely small $T$) we observe in the stationary state the $X^{C}$ 
Nash configuration with perhaps few different strategies. The above argument was made 
into a rigorous proof for an infinite system of the closely related lattice-gas model 
(the Blume-Capel model) of interacting particles by Bricmont and Slawny in \cite{brsl1}.  

In the above example, $X^{B}$ and $X^{C}$ 
have the same total payoff but $X^{C}$ has more lowest-cost excitations
and therefore it is low-noise ensemble stable. We will now discuss the situation,
where $X^{C}$ has a smaller total payoff but nevertheless in the long run
$C$ is played with a frequency close to $1$ if the noise level is low but not extremely low. 
We will consider a family of games with the following payoff matrix:
\vspace{2mm}

\noindent {\bf Example 4}
\vspace{2mm}

\hspace{25mm} A  \hspace{3mm} B \hspace{3mm} C  

\hspace{15mm} A \hspace {3mm} 1.5  \hspace{3mm} 0 \hspace{5mm} 1

U = \hspace{6mm} B \hspace{3mm} 0  \hspace{3mm} $2+\alpha$ \hspace{2mm} 1

\hspace{15mm} C \hspace{3mm} 1  \hspace{6mm} 1 \hspace{4mm} 2,

where $\alpha>0$ so $B$ is both payoff and pairwise risk-dominant.

We are interested in the long-run behavior of our system for small positive $\alpha$
and low $T$. One may modify the proof of Theorem 1 (see Appendix B) and obtain 
the following theorem.
\begin{theo}
For every $\epsilon>0$, there exist $\alpha(\epsilon)$ and $T(\epsilon)$ 
such that for every $0<\alpha<\alpha(\epsilon)$, 
there exists $T(\alpha)$ such that for $T(\alpha)<T<T(\epsilon)$, 
$X^{C}$ is $\epsilon$-ensemble stable,
and for $0<T<T(\alpha)$, $X^{B}$ is $\epsilon$-ensemble stable.
\end{theo}
Observe that for $\alpha=0$, both $X^{B}$ and $X^{C}$ are stochastically stable
(they appear with the frequency $1/2$ in the limit of zero noise) but $X^{C}$
is low-noise ensemble stable. For small $\alpha > 0$, $X^{B}$ is both stochastically
(it appears with the frequency $1$ in the limit of zero noise) and low-noise 
ensemble stable. However, for intermediate noise 
$T(\alpha)<T<T(\epsilon)$, if the number of players is big enough, then in the long run, 
almost all players use the strategy $C$ - $X^{C}$ is ensemble stable). 
If we lower T below $T(\alpha)$, then almost all players start to use the strategy $B$. 
$T=T(\alpha)$ is the line of the first-order phase transition. In the thermodynamic limit, 
there exist two Gibbs state (equilibrium states) on this line. We may say that at $T=T(\alpha)$, 
the society of players undergoes a {\em phase transition} from $C$ to $B$-behavior. 
\vspace{1mm}

Now we will consider games with a dominated strategy and two symmetric Nash equilibria. 
We say that a given strategy is is {\em dominated} if it gives a player 
the lowest payoff regardless of a strategy chosen by an opponent.
It is easy to see that dominated strategies cannot be present in any Nash equilibrium. 
Therefore such strategies should not be used by players and consequently we might think 
that their presence should not have any impact on the long-run behavior of the system. 
We will show in the following example that this may not be necessarily true.
\eject

\noindent {\bf Example 5}
\vspace{2mm}

\hspace{23mm} A  \hspace{5mm} B \hspace{5mm} C  

\hspace{15mm} A \hspace {3mm} 0  \hspace{5mm} 0.1 \hspace{4mm} 1

U = \hspace{6mm} B \hspace{3mm} 0.1  \hspace{1mm} $2+\alpha$ \hspace{1mm} 1

\hspace{15mm} C \hspace{3mm} 1  \hspace{5mm} 1 \hspace{6mm} 2,

where $\alpha>0$.
\vspace{3mm}

We see that strategy $A$ is dominated by $B$ and $C$ 
hence $X^{A}$ is not a Nash equilibrium. $X^{B}$ and $X^{C}$
are both Nash equilibria but only $X^{B}$ is a ground-state 
configuration for $-U.$ In the absence of $A$,
$B$ is both payoff and risk-dominant and therefore is stochastically stable
and low-noise ensemble stable. Adding the strategy $A$ does not change dominance 
relations; $B$ is still payoff and pairwise risk dominant.
However, we may modify slightly the proof of Theorem 2 to show that 
$X^{C}$ is $\epsilon$-ensemble stable at intermediate noise levels. 
The mere presence of the dominated strategy $A$ changes the long-run behavior of the system. 
Similar results were already discussed in adaptive games of Myatt and Wallace \cite{wallace}. 
In their games, at every discrete moment of time, one of the agents leaves 
the population and is replaced by another one who plays the best response. 
He calculates his best response with respect to his own payoff matrix 
which is the matrix of a common average payoff disturbed by a realization 
of some random variable with the zero mean. The noise does not appear in the game 
as a result of players' mistakes but is the effect of their idiosyncratic preferences. 
The authors then show that the presence of a dominated strategy may change 
the stochastic stability of equilibria. However, the reason for such a behavior
is different in their and in our models. In our model, it is relatively easy 
to get out of $X^{C}$ and this makes $X^{C}$-ensemble stable. Mayatt and Wallace introduce
a dominated strategy in such a way that it is relatively easy to make a transition to it 
from a risk and payoff-dominant configuration and then with a high probability 
the system moves to a third Nash configuration which results in its stochastic stability.

Although in above models, the number of players was very large, 
their strategic interactions were decomposed into a sum of two-player games. 
Stochastic and ensemble stability of three-player games were 
investigated in \cite{physica}.

\section{Stochastic Stability in Non-potential Games}

Let us now consider games with three strategies and three
symmetric Nash equilibria: $(A,A), (B,B)$, and $(C,C)$. 
Generically, such games do not have a potential and therefore their stationary 
states cannot be explicitly constructed. To find them, we must resort to different methods. 
We will use a tree representation of the stationary distribution of Markov chains 
\cite{freiwen1,freiwen2} (see also Appendix C). 

To illustrate this technique we will discuss a following two-player game 
with three strategies. 
\vspace{2mm}

\noindent {\bf Example 6}
\vspace{2mm}

\noindent Players are located on a finite subset $\Lambda$ of ${\bf Z}$ 
(with periodic boundary conditions) and interact with their 
two nearest neighbors. They have at their disposal three pure strategies: 
$A, B,$ and $C$. The payoffs are given by the following matrix:
\vspace{5mm}

\hspace{21mm} A  \hspace{2mm} B \hspace{2mm} C  

\hspace{15mm} A  \hspace{2mm} 3  \hspace{2mm} 0 \hspace{2mm} 2

U = \hspace{7mm} B \hspace{2mm} 2  \hspace{2mm} 2 \hspace{2mm} 0

\hspace{15mm} C \hspace{2mm} 0  \hspace{2mm} 0 \hspace{2mm} 3

Our game has three Nash equilibria, $(A,A), (B,B)$, and $(C,C)$.
Let us note that in pairwise comparisons, $B$ risk dominates $A$, 
$C$ dominates $B$ and $A$ dominates $B$. The corresponding spatial 
game has three homogeneous Nash configurations: $X^{A}, X^{B}$, and $X^{C}$. 
They are the only absorbing states of the noise-free best-response dynamics. 
When we start with any state different from $X^{A}$, $X^{B}$, and $X^{C}$, 
then after a finite number of steps we arrive at either $X^{A}$, $X^{B}$ or $X^{C}$ 
and then stay there forever. It follows from the tree representation of stationary 
states (see Appendix C) that any state different from $X^{A}$, $X^{B}$, and $X^{C}$, 
has zero probability in the stationary distribution in the zero-noise limit. 
Moreover, in order to study the zero-noise limit of the stationary distribution, 
it is enough to consider probabilities of transitions between absorbing states. 

\begin{theo}
$X^{B}$ is stochastically stable
\end{theo} 

{\bf Proof}: The following are maximal A-tree, B-tree, and C-tree:
$$B \rightarrow C \rightarrow A, \hspace{3mm} C \rightarrow A \rightarrow B,
\hspace{3mm} A \rightarrow B \rightarrow C.$$

Let us observe that
\begin{equation}
P_{B \rightarrow C \rightarrow A}= O(e^{-6/T}),
\end{equation}
\begin{equation}
P_{C \rightarrow A \rightarrow B}= O(e^{-4/T}),
\end{equation}
\begin{equation}
P_{A \rightarrow B \rightarrow C}= O(e^{-6/T}),
\end{equation}
where $\lim_{x \rightarrow 0}O(x)/x =1.$

The theorem follows from the tree characterization of stationary states
as described in Appendix C.

$X^{B}$ is stochastically stable because it is much more probable (for low $T$) 
to escape from $X^{A}$ and $X^{C}$ than from $X^{B}$. The relative payoffs 
of Nash configurations are not relevant here (in fact $X^{B}$ has the smallest payoff).   
Let us recall Example 3 of a potential game, where an ensemble-stable configuration
has more lowest-cost excitations. It is easier to escape 
from an ensemble-stable configuration than from other Nash configurations. 

Stochatic stability concerns single configurations in the zero-noise limit; 
ensemble stability concerns families of configurations 
in the limit of the infinite number of players. It is very important 
to investigate and comparethese two concepts of stability in nonpotential games.  

Nonpotential spatial games cannot be directly presented as systems of interacting
particles. They constitute a large family of interacting objects not thoroughly studied 
so far by methods statistical physics. Some partial results concerning stochastic stability 
of Nash equilibria in nonpotential spatial games were obtained in 
\cite{ellis1,ellis2,blume1,physica,statphys}.

One may wish to say that $A$ risk dominates the other two strategies 
if it risk dominates them in pairwise comparisons. In Example 6, 
that $B$ dominates $A$, $C$ dominates $B$, and finally $A$ dominates $C$. 
But even if we do not have such a cyclic relation of dominance, 
a strategy which is pairwise risk-dominant may not be stochastically stable \cite{statphys}. 
A more relevant notion seems to be that of a global risk dominance \cite{mar}.
We say that $A$ is globally risk dominant if it is a best response
to a mixed strategy which assigns probability $1/2$ to $A$.
It was shown in \cite{ellis1,ellis2} that a global risk-dominant 
strategy is stochastically stable in some spatial games with local interactions. 

A different criterion for stochastic stability was developed by Blume
\cite{blume1}. He showed (using techniques of statistical mechanics) 
that in a game with $k$ strategies $A_{i}$ and $k$ symmetric Nash equilibria
$(A_{i},A_{i})$, $i=1,...,k$ 
and $k$ pure symmetric Nash equlibria, $A_{1}$ is stochastically stable if
\begin{equation}
\min_{n>1}(U(A_{1},A_{1})-U(A_{n},A_{1})) > \max_{n>1}(U(A_{n},A_{n})-U(A_{1},A_{n})).
\end{equation}
We may observe that if $A_{1}$ satisfies the above condition, then it is pairwise
risk dominant. 

\section{Discussion}

We discussed effects of the number of players and the noise level 
on the long-run behavior in the stochastic dynamics of spatial games. 
In the so-called potential games with the Boltzmann-type updating, stationary states 
are Gibbs distributions of corresponding lattice-gas models.
We used ideas and techniques of statistical mechanics to analyze such games.
   
In particular, we were concerned with two limits of our models. 
In the first one, for a fixed number of players, one considers an arbitrarily 
low level of noise. Then the relevant concept is that of stochastic stability 
of single configurations. For a fixed level of noise,
in the limit of the infinite number of players, long-run behavior is described
by the stability of certain ensembles of configurations. We show in several examples that 
the long-run behavior may be different in these two limiting cases. 

In non-potential games, stationary states cannot be explicitly constructed as before.
In order to study their zero-noise limits, one may use their tree representation. 
We illustrated this technique on a simple example. Constructing stationary states 
in non-potential spatial games is an important open problem.
\vspace{3mm}

\noindent {\bf Acknowledgments}: I thank Christian Maes and Joseph Slawny
for useful conversations. Financial support by the Polish Committee 
for Scientific Research under the grant KBN 5 P03A 025 20 is kindly acknowledged.

\appendix
\section{}
Here we provide a proof of (4.2). 
We follow \cite{brsl1} very closely.
We begin by defining formally restricted ensembles. Let
$$\Omega=\{A,B,C\}^{Z^{2}}$$
be the configuration space of our model. 
$$\Omega_{R}^{B}=\{X  \in  \Omega,  X_{i}=B, C \; for  \;  all  \; i \in Z^{2} \; 
and \; if \; X_{i}=C, \;  then \; X_{j}=B \; if \; |i-j|=1\},$$
$$\Omega_{R}^{C}=\{X  \in  \Omega, \; \; if X_{i}=A \; or \; B, \; then \; X_{j}=C 
\; if \; |i-j|=1\}$$
are the restricted ensembles of configurations of the lowest-cost excitations 
of $X^{B}$ and $X^{C}$ Nash configurations. Observe that $X^{C}$
has many more lowest-cost excitations than $X^{B}$. 

We define partition functions of restricted ensembles with boundary conditions 
$Y\in\Omega_{R,\Lambda^{c}}^{k},$ $k=B,C$ as
\begin{equation}
Z_{R}(\Lambda|Y)=\sum e^{\beta U_{\Lambda}(X)},
\end{equation}
where the sum is over $X \in \Omega_{R}^{k}$ which are equal to $Y$ on $\Lambda^{c}$, 
\begin{equation}
U_{\Lambda}(X)=\sum_{\{i,j\}\cap \Lambda \neq \emptyset} U(X_{i},X_{j}),
\end{equation}
and $\beta=1/T$.
It is a standard result in rigorous statistical mechanics that
a following limit exists
\begin{equation}
\psi_{R}(\beta|k)=\lim_{\Lambda \rightarrow Z^{2}}log \frac{Z_{R}(\Lambda|Y)}{|\Lambda| \beta}
\end{equation} 
and has a convergent expansion. $\psi_{R}(\beta|k)$ is called a thermodynamic potential of a gas 
of noninteracting lowest-cost excitations. We may write
\begin{equation}
logZ_{R}(\Lambda|Y)=|\Lambda|\beta\psi_{R}(\beta|k)+o(e^{-4\beta})|\delta\Lambda|, k=B,C,
\end{equation}
where
\begin{equation}
\beta\psi_{R}(\beta|B)=2+e^{-4\beta} +O(e^{-8\beta}), 
\end{equation}
\begin{equation}
\beta\psi_{R}(\beta|C)=2+2e^{-4\beta} +O(e^{-8\beta}).
\end{equation}
and $\delta\Lambda$ is the boundary of $\Lambda$.

We define $ret(X)$ by $ret(X)_{i}=B$ if  $X_{i}=C$ and  $X_{j}=B$ for $|i-j|=1$, 
$ret(X)_{i}=C$ if  $X_{i}=A,B$ and  $X_{j}=C$ for $|i-j|=1$, 
and $ret(X)_{i}=X_{i}$ otherwise. Therefore, in $ret(X)$ we remove all lowest-cost excitations 
of $X$ but not excitations of a higher cost. If $X \in \Omega_{R}^{B}$ ($\Omega_{R}^{C}$),
then $ret(X)=X^{B}$ ($X^{C})$. Let us define the boundary of $X$ as 
the set of pairs $(i,j)$  such that $ret(X)_{i} \neq ret(X)_{j}$  
A small scale contour $\gamma$ of a configuration $X$ is a pair
$\gamma=([\gamma],X_{[\gamma]}),$ where $[\gamma]$ is the maximal connected subset 
of the union of the boundary of $X$ and pairs of sites $(i,j)$ 
such that $X_{i}=X_{j}=A.$ The cost of $\gamma$ is
$$U(\gamma)=\sum_{(i,j) \in \gamma} (2-U(X_{i},X_{j}))$$
Now we define large-scale contours. Let $L(\beta)=e^{5 \beta/2}.$
We cover $Z^{2}$ with squares
$$B(i)=B(o)+(1/2)Li, i \in  Z^{2},$$
where $B(o)$ is the square of side $L(\beta)$ centered at the origin and containing $e^{5\beta}$
lattice sites. We call $B(i)$ a {\em regular} box of $X$ if $ X_{B(i)} \in \Omega_{R, B(i)}^{C}$
and it is {\em irregular} otherwise. There are two types of irregular boxes of $X$:

type 1 \hspace{3mm} if  $X_{B(i)} \in \Omega_{R,B(i)}^{B},$

type 2 \hspace{3mm} if  a small-scale contour of $X$  intersects $B(i).$

A large-scale contour $\Gamma$ is a connected family of irregular squares.
Let $||\Gamma||$ be the number of squares in $\Gamma$ and $|\Gamma|$ 
the number of lattice sites in $\Gamma,$  $[\Gamma]=\cup_{B \in \Gamma}B.$
For any function $f$ on $\Omega$
\begin{equation}
P_{\Lambda}(f|Y)=\sum f(X)
\frac{e^{\beta \sum_{\{i,j\}\cap \Lambda \neq \emptyset}U(X_{i},X_{j})}}{Z(\Lambda|Y)},
\end{equation}
where the sum is over $X \in \Omega$ which are equal to $Y$ on $\Lambda^{c}.$
For $[\Gamma] \subset \Lambda$, let $P_{\Lambda}(\Gamma|Y)=P_{\Lambda}(\chi_{\Lambda}|Y)$,
where $\chi_{\Lambda}(X)=1$ if $\Gamma$ is a contour of $X$ and zero otherwise.
Therefore 
\begin{equation}
P_{\Lambda}(\Gamma|Y)=\frac{Z(\Lambda|\Gamma,Y)}{Z(\Lambda|Y)},
\end{equation}
where 
\begin{equation}
Z(\Lambda|\Gamma,Y)=\sum e^{\beta U_{\Lambda}(X)},
\end{equation}
where the sum is over $X \in \Omega$ which are equal to $Y$ on $\Lambda^{c}$
and contain $\Gamma$.
$P_{\Gamma}(\bullet|Y)$ is called a Gibbs measure in $\Lambda$ with boundary conditions $Y$.
Now we are ready to formulate our main proposition.
\begin{prop}
For big enough $\beta$ there exists $c$ such that for all finite $\Lambda \subset Z^{2}$, 
all boundary conditions
$Y \in  \Omega_{R,\Lambda^{c}}^{C}$ and all contours $\Gamma$ contained in $\Lambda$

$P_{\Lambda}(\Gamma|Y) \leq e^{-c\beta||\Gamma||}.$

\end{prop}

{\bf Proof}: First we condition on strategies in $\delta[\Gamma]$,
\begin{equation}
P_{\Lambda}(\Gamma|Y)=\sum_{Z}P_{\Lambda}(\Gamma|Y,Z)P_{\Lambda}(Z|Y).
\end{equation}
Then we get

\begin{equation}
P_{\Lambda}(\Gamma|Y,Z)=P_{[\Gamma]}(\Gamma|Y,Z)=\frac{Z([\Gamma]|\Gamma,Y,Z)}{Z([\Gamma]|Y,Z)},
\end{equation}

\begin{equation}
Z([\Gamma]|\Gamma,Y,Z)=\sum_{\Gamma^{2}}\sum_{\omega}Z([\Gamma]|\Gamma^{2},\omega,Y,Z),
\end{equation}
where the first summation is over all possible families $\Gamma^{2}$ of type-2 squares of $\Gamma$
and the second over families  $\omega$ of small-scale contours in $[\Gamma]$ 
such that for each square of $\Gamma^{2}$ there is a contour of $\omega$ intersecting the square. 
Let
$$[\Gamma]- [\omega]=\cup_{a} M_{a}, \; [\omega]=\cup_{\gamma \in \omega}[\gamma]$$
be the decomposition of $[\Gamma]-[\omega]$ into connected components. 
Now we have
\begin{equation}
Z([\Gamma]|\Gamma^{2},\omega , Y,Z)= e^{2\beta\sum_{\gamma \in \omega}|\gamma|} e^{-\beta
U(\omega)}\Pi_{a}Z_{R}(M_{a}|X_{a}),
\end{equation}
where
\begin{equation}
U(\omega)=\sum_{\gamma \in \omega}U(\gamma),
\end{equation}
$|\gamma|$ is the number of pairs in $\gamma$ and $X_{a}$ is the configuration on $\delta M_{a}$.

After inserting (A.13) into (A.12) and (A.12) into (A.11) we have to estimate the ratio
\begin{equation}
\frac{(\Pi_{a}Z_{R}(M_{a}|X_{a}))}{Z([\Gamma]|Y,Z)} 
\leq \frac{(\Pi_{a}Z_{R}(M_{a}|X_{a}))}{Z_{R}([\Gamma] |Y,Z)},
\end{equation}
where in the dominator we used the lower bound
\begin{equation}
Z([\Gamma]|Y,Z) \geq   Z_{R}([\Gamma]|Y,Z).
\end{equation}

We write the volume terms of (A.15) as
\begin{equation}
e^{\beta(\sum_{a}|M_{a}|\psi_{R}(\beta|k(a))-|\Gamma|\psi_{R}(\beta|C))}
\leq e^{-(e^{-4\beta}+O(e^{-8\beta}))\sum_{a: k(a) \neq C}|M_{a}|}
\end{equation}
$$\leq e^{-(1/2)|\Gamma^{1}| e^{-4\beta}} = e^{-(1/2)||\Gamma^{1}|| e^{\beta}},$$
where $\Gamma^{1}=\Gamma-\Gamma^{2}.$
We also have to estimate boundary terms. The family of boundaries of $\delta M_{a}$
consist of two subfamilies: 
one contained in $[\omega]$ and another contained in $\delta[\Gamma]$, 
on which we have the same boundary conditions, 
$Y$ and $Z$, in the numerator and the denominator of (A.15). 
Since these boundary conditions are the same, contributions to the boundary term 
cancel each other. Finally using  $U(\gamma)> |\gamma|$
we obtain that (A.15) is bounded by
\begin{equation}
e^{-(1/2)||\Gamma^{1}||e^{\beta}+c'|\omega|e^{-4\beta}}
\end{equation} 
Therefore
\begin{equation}
\frac{Z([\Gamma]|\Gamma^{2},\omega, Y,Z)}{Z_{R}([\Gamma]|Y,Z)} 
\leq e^{-\beta'U(\omega)-(1/2)||\Gamma^{1}|| e^{\beta}},
\end{equation}
where 
\begin{equation}
\beta'=\beta-c'e^{-4\beta}.
\end{equation}
We obtain that
\begin{equation}
P_{\Lambda}(\Gamma|Y,Z) \leq e^{-(1/2)||\Gamma^{1}||e^{\beta}}\sum_{\omega}
e^{-\beta'U(\omega)},
\end{equation}
where the sum over the families $\omega$ of small scale-contours is restricted 
by the condition that for each $B \in \Gamma$ there exists at least one contour 
$\gamma \in \omega$ with $[\gamma] \cap B \neq \emptyset.$
We get
\begin{equation}
\sum_{\omega}e^{-\beta'U(\omega)} \leq \Pi_{B \in \Gamma^{2}} 
(\sum_{m \geq 1}(1/m!)\sum^{B}_{\gamma_{1},...,\gamma_{m}}e^{-\beta'\sum_{j}U(\gamma_{j})}
\end{equation}
$$ \leq \Pi_{B \in \Gamma^{2}} (\sum_{m \geq 1}(1/m!)(\sum^{B}_{\gamma}
e^{-\beta'U(\gamma)})^{m}),$$
where the superscript $B$ indicates summation over contours $\gamma$ with $[\gamma] 
\cap B \neq \emptyset$

Now because $U(\gamma) \geq |\gamma|$ and $U(\gamma) \geq 6$, for big $\beta$ we get
\begin{equation}
\sum^{B}_{\gamma}e^{-\beta'U(\gamma)} \leq c''|B|e^{-6\beta'}=c''e^{-\beta}.
\end{equation}
From (A.22) and (A.23) we get
\begin{equation}
(e^{c''e^{-\beta}}-1)||\Gamma^{2}|| \leq(c'''e^{-\beta})^{||\Gamma^{2}}.
\end{equation} 
We conclude the proof by using the above estimate in (A.21).

Now the following proposition is a consequence of Proposition 1
\begin{prop}
There exist two positive constants, $c$ and $c'$, such that
$P_{\Lambda}(||\Gamma||>c|\delta \Lambda| \; |Y) \leq e^{-c'\beta|\delta \Lambda|}$
for big enough $\beta$ and for all finite $\Lambda \subset Z^{2}$, $\Gamma$ in $\Lambda$, 
and all boundary conditions $Y \in  \Omega_{\Lambda}^{c}$. 
\end{prop}

{\bf Proof}: We change boundary conditions from an arbitrary $Y$ to $C$. We have
\begin{equation}
P_{\Lambda}(\bullet|Y) \leq e^{4\beta|\delta \Lambda|}P_{\Lambda}(\bullet|C).
\end{equation}

We connect disconnected parts of $\Gamma$ through $\delta\Lambda$ 
and from Proposition 1 we get

\begin{equation}
P_{\Lambda}(||\Gamma||>c|\delta \Lambda| \; |C) 
\leq e^{-c'\beta|\delta \Lambda|}
\end{equation}

which finishes our proof.

{\bf Proof of (4.2)}:

By Proposition 2 we may assume that $\Gamma$
covers a small part of $\Lambda$. Indeed, with high probability we have
\begin{equation}
|\Gamma| = e^{5\beta}||\Gamma|| \leq O(e^{5\beta})|\delta \Lambda|.
\end{equation}

In the complement of $[\Gamma]$
we have the gas of noninteracting lowest-cost excitations of $X^{C}$
which are very rare if $\beta$ is large enough so the noise level $T=1/\beta$
is low enough. This proves that there is the unique limit
$\lim_{\Lambda \rightarrow Z^{2}}P_{\Lambda}(\bullet|Y)$ which is equal to $\mu^{T}$
in (4.1) and (4.2) is established. 

\section{}

The payoff of $X^{B}$ in Example 5 is bigger than that of $X^{C}.$
However, for small $\alpha$, $X^{C}$ has again larger thermodynamic potential.
Thermodynamic potentials of lowest-cost excitations have following expansions:
\begin{equation}
\beta\psi_{R}(\beta|B)=2 + \alpha + e^{-4(1+\alpha)\beta} + O(e^{-8(1+\alpha)\beta}).
\end{equation}
\begin{equation}
\beta\psi_{R}(\beta|C)=2 + 2e^{-4\beta} + O(e^{-8\beta}).
\end{equation}

If $\alpha < \frac{1}{2}e^{-4\beta}$, then
\begin{equation}
\beta(\psi_{R}(\beta|C)-\psi_{R}(\beta|B)) > \frac{1}{2}e^{-4\beta}.
\end{equation}

Now to prove Theorem 2 we may repeat the proof of Theorem 1.

\section{}

\noindent The following tree representation of stationary states 
of Markov chains was proposed by Freidlin and Wentzell (1970 and 1984). 
Let $(\Omega,P)$ be an irreducible Markov chain with a state space 
$\Omega$ and transition probabilities given by $P: \Omega \times \Omega \rightarrow [0,1]$. 
It has a unique stationary probability distribution $\mu$ (called also a stationary state). 
For $X \in \Omega$, an $X$-tree is a directed graph 
on $\Omega$ such that from every $Y \neq X$ there is a unique path to $X$
and there are no outcoming edges out of $X$. Denote by $T(X)$ the set of all $X$-trees
and let 
\begin{equation}
q(X)=\sum_{d \in T(X)} \prod_{(Y,Y') \in d}P(Y,Y'),
\end{equation}
where the product is with respect to all edges of $d$. 
Now one can show that
\begin{equation}
\mu(X)=\frac{q(X)}{\sum_{Y \in \Omega}q(Y)}
\end{equation}
for all $X \in \Omega.$

In our case, $P$ is given by (3.1). A state is an absorbing one 
if it attracts nearby states in the noise-free best-response dynamics. 
Let us assume that after a finite number of steps of the noise-free dynamics 
we arrive at one of the absorbing states (there are no other recurrence classes)
and stay there forever. Then it follows from the above tree representation 
that any state different from absorbing states has zero probability 
in the stationary distribution in the zero-noise limit. 
Moreover, in order to study the zero-noise limit of the stationary state, 
it is enough to consider paths between absorbing states. More precisely, 
we construct $X$-trees with absorbing states as vertices; the family of such 
$X$-trees is denoted by $\tilde{T}(X)$. Let 
\begin{equation}
q_{m}(X)=max_{d \in \tilde{T}(X)} \prod_{(Y,Y') \in d}\tilde{P}(Y,Y'),
\end{equation}
where $\tilde{P}(Y,Y')= max \prod_{(W,W')}P(W,W')$,
where the product is taken along any path joining $Y$ with $Y'$ and the maximum 
is taken with respect to all such paths. 
Now we may observe that if $lim_{\epsilon \rightarrow 0} q_{m}(Y)/q_{m}(X)=0,$
for any $Y\neq X$, then $X$ is stochastically stable. Therefore we have to compare  
trees with the biggest products in (C.3); such trees we call maximal.


\begin{thebibliography}{99}

\bibitem{santa}{{\em The Economy as an Evolving Complex System II}, 
Arthur W B, Durlauf S N, and Lane D A, eds. 1997 (Addison-Wesley, Reading MA)}
\bibitem{young2}{Young P H 1998 {\em Individual Strategy and Social
Structure: An Evolutionary Theory of Institutions} (Princeton University Press, Princeton)}
\bibitem{young3}{{\em Social Dynamics}, Durlauf S N and Young P H, eds. 1998 (MIT Press, Cambridge MA)}
\bibitem{nowsig}{Nowak M A and Sigmund K 2004 {\em Science} {\bf 303} 793}
\bibitem{econo}{Econophysics bulletin on www.unifr.ch/econophysics}

\bibitem{wei}{Weibull J 1995 {\em Evolutionary Game Theory} 
(MIT Press, Cambridge, MA)}
\bibitem{hof2}{Hofbauer J and Sigmund K 1998 {\em Evolutionary Games and Population Dynamics}
(Cambridge University Press, Cambridge)}
\bibitem{ams}{Hofbauer J and Sigmund K 2003  
{\em Bulletin AMS} {\bf 40} 479} 

\bibitem{blume1}{Blume L E 1993 {\em Games Econ. Behav.} {\bf 5} 387} 
\bibitem{ellis1}{Ellison G 1993 {\em Econometrica} {\bf 61} 1047} 
\bibitem{ellis2}{Ellison G 2000 {\em Review of Economic Studies} {\bf 67} 17}
\bibitem{nowak1}{Nowak M A and May R M 1993  
{\em Int. J. Bifurcation and Chaos} {\bf 3} 35}
\bibitem{nowak2}{Nowak M A, Bonhoeffer S, and May R M 1994 
{\em Int. J. Bifurcation and Chaos} {\bf 4} 33}
\bibitem{linnor}{Lindgren K and Nordahl M G 1994
{\em Physica D} {\bf 75} 292} 
\bibitem{doebeli}{Brauchli K, Killingback T, and Doebeli M 1999
{\em J. Theor. Biol.} {\bf 200}: 405}
\bibitem{sabo}{Szab\'{o} G, Antal T, Szab\'{o} P, and Droz M 2000 
{\em Phys. Rev. E} {\bf 62} 1095}
\bibitem{hauert}{Hauert Ch 2002 {\em Int. J. Bifurcation and Chaos} {\bf 12} 1531} 

\bibitem{foya}{Foster D and Young P H {\em Theoretical Population Biology} {\bf 38} 219} 
\bibitem{mon}{Monderer D and Shapley L S 1996 {\em Games Econ. Behav.} {\bf 14} 124}
\bibitem{freiwen1}{Freidlin M and Wentzell A 1970 {\em Russian Math. Surveys}
{\bf 25} 1}
\bibitem{freiwen2}{Freidlin M and Wentzell A 1984
{\em Random Perturbations of Dynamical Systems} (Springer Verlag, New York).}

\bibitem{hs}{Hars\'{a}nyi J and Selten R 1988 {\em A General Theory of Equilibrium
Selection in Games} (MIT Press, Cambridge)} 

\bibitem{brsl1}{Bricmont J and Slawny J 1986  {\em First order phase transitions 
and perturbation theory in Statistical Mechanics and Field Theory: Mathematical Aspects}
(Lecture Notes in Physics 257. Springer-Verlag} 
\bibitem{brsl2}{Bricmont J and Slawny J 1989 {\em J. Stat. Phys.} {\bf 54} 89}
\bibitem{wallace}{Myatt D P and Wallace C 2003 {\em J. Econ. Theory} {\bf 113} 286}

\bibitem{physica}{Mi\c{e}kisz J 2004 {\em Stochastic stability of spatial three-player games.}
Warsaw University preprint, www.mimuw.edu.pl/$\sim$miekisz/physica.ps, 
to appear in {\em Physica A}}
\bibitem{statphys}{Mi\c{e}kisz J 2004. {\em Stochastic stability in spatial games.}
Warsaw University preprint, www.mimuw.edu.pl/$\sim$miekisz/statphys.ps, to appear in {\em J. Stat. Phys.}} 

\bibitem{mar}{Maruta T 1997 {\em Games Econ. Behav.} {\bf 19} 2211} 

\end{thebibliography}
\end{document}